\documentclass[aps,prl,reprint,superscriptaddress]{revtex4-1}

\usepackage{graphicx}
\usepackage{dcolumn}
\usepackage{bm}

\begin{document}


\title{Superconducting nanowires by electron-beam-induced deposition}



\author{Shamashis Sengupta}
\email[]{shamashis.sengupta@u-psud.fr}
\affiliation{Centre de Sciences Nucleaire et de Sciences de la Matiere, UMR 8609, CNRS, Univ. Paris-Sud, 91405 ORSAY Cedex, France}
\affiliation{Laboratoire de Physique des Solides, UMR 8502, CNRS, Univ. Paris-Sud, 91405 ORSAY Cedex, France}
\author{Chuan Li}
\affiliation{Laboratoire de Physique des Solides, UMR 8502, CNRS, Univ. Paris-Sud, 91405 ORSAY Cedex, France}
\author{Cedric Baumier}
\affiliation{Centre de Sciences Nucleaire et de Sciences de la Matiere, UMR 8609, CNRS, Univ. Paris-Sud, 91405 ORSAY Cedex, France}
\author{Alik Kasumov}
\affiliation{Laboratoire de Physique des Solides, UMR 8502, CNRS, Univ. Paris-Sud, 91405 ORSAY Cedex, France}
\affiliation{Institute of Microelectronics Technology and High Purity Materials, RAS, ac. Ossipyan, 6, Chernogolovka, Moscow Region, 142432, Russia}
\author{S. Gueron}
\affiliation{Laboratoire de Physique des Solides, UMR 8502, CNRS, Univ. Paris-Sud, 91405 ORSAY Cedex, France}
\author{H. Bouchiat}
\affiliation{Laboratoire de Physique des Solides, UMR 8502, CNRS, Univ. Paris-Sud, 91405 ORSAY Cedex, France}
\author{F. Fortuna}
\affiliation{Centre de Sciences Nucleaire et de Sciences de la Matiere, UMR 8609, CNRS, Univ. Paris-Sud, 91405 ORSAY Cedex, France}


\date{\today}

\begin{abstract}
Superconducting nanowires can be fabricated by decomposition of an organometallic gas using a focused beam of Ga ions. However, physical damage and unintentional doping often results from the exposure to the ion beam, motivating the search for a means to achieve similar structures with a beam of electrons instead of ions. This has so far remained an experimental challenge. We report the fabrication of superconducting tungsten nanowires by electron-beam-induced-deposition, with critical temperature of 2.0 K and critical magnetic field of 3.7 T, and compare them with superconducting wires made with ions. This work opens up new possibilities for the realization of nanoscale superconducting devices, without the requirement of an ion beam column. 

\end{abstract}


\maketitle

Synthesis of superconducting nanowires by direct writing with the electron beam can have a significant impact on fabrication of nanoscale circuits for contacting, repairing and producing devices and in research areas like mesoscopic physics and applied superconductivity. But previous attempts in this direction have been unsuccessful. It is known that a metallic nanowire can be fabricated by decomposing an organometallic gas by focusing a beam of Ga ions, \cite{melngailis, matsui, langford, utke} and experiments \cite{sadki} have demonstrated tungsten nanowires made with such ion-beam-induced deposition (IBID) to be superconducting with a high critical field (9.5 T) and temperature (5.2 K) for practical realization in cryogenic experiments. IBID of tungsten has found applications in Josephson junction circuits \cite{chiodi, huth_josephson}, low-temperature transport in molecules \cite{kasumov}, study of superconductivity in ferromagnetic nanowires \cite{wang_chan}, experiments on the vortex glass state \cite{chan} and observation of the vortex lattice \cite{guillamon1} and vortex melting transition \cite{guillamon2}.  However, irradiation by the FIB can cause considerable damage to the sample due to etching and implantation of the energetic Ga ions, this being more acute in the case of nanomaterials. \cite{nam} In the case of semiconducting nanowires (in particular InAs), diffusion of Ga has been seen to modify the intrinsic electronic properties of the material itself. \cite{kasumov_unpub} It is desirable to develop a deposition method less destructive, specially keeping in mind potential application to materials that are thin and more vulnerable to mechanical damage by the FIB, e.g. nanotubes, systems like graphene, ultra-thin layers of MoS$_2$ and WS$_2$ - whose unique properties in the atomically thin form have become an extremely active field of research. In addition, such a fabrication method with an electron beam \cite{juntao, mitsuishi, dorp, hagen, bishop} would also be less expensive since it does not require a FIB source and only a scanning electron microscope (SEM) fitted with a gas injection system (for supplying the precursor) is sufficient. However, it has proved challenging in the past to obtain nanowires by EBID that retain even their metallic properties at low temperatures, not to mention superconductivity. \cite{ebeam1, ebeam2} We report here the optimization of electron-beam-induced deposition of tungsten nanowires which are not only good conductors at low temperatures, but observed to be superconducting as well. To the best of our knowledge, this is the first work to report superconductivity in EBID structures.

\begin{figure}
\begin{center}
\includegraphics[width=70mm]{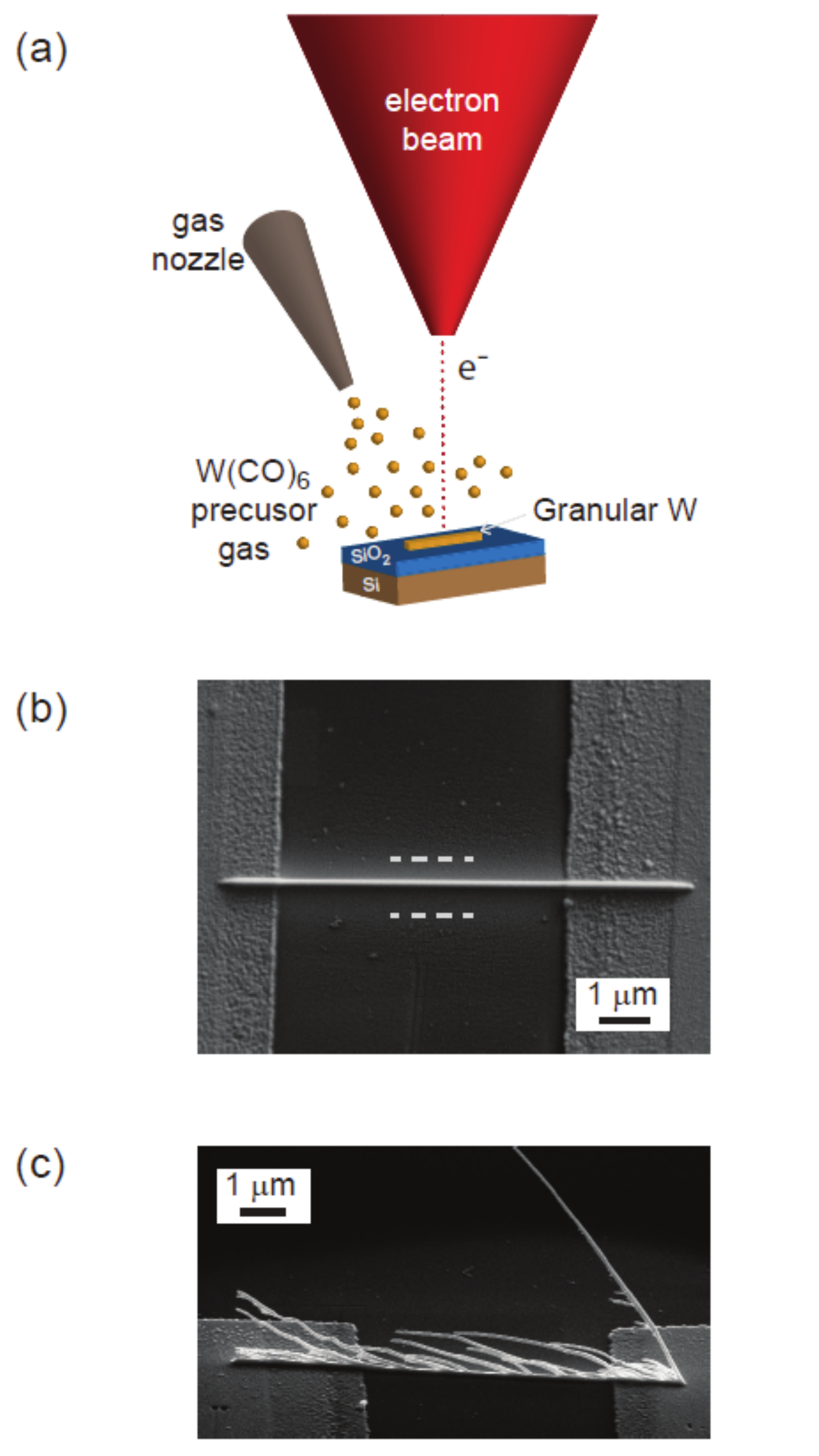}
\caption{(Color online) (a) A schematic of the setup for fabricating tungsten wires with electron-beam-induced deposition (EBID). (b) A tungsten wire obtained by EBID with the following parameters: 10 keV electron energy, 5.1 nA beam current, 1 nm step size, 10 ms dwell time and 10 times pattern repeat. (c) `Pillar-like' growth of tungsten wires was observed when the dwell time was increased to 100 ms.}
\end{center}
\end{figure}

EBID is carried out in the chamber of a scanning electron microscope (SEM). Our dual beam system is also equipped with a focused ion beam (FIB) source for Ga ions. Either EBID or IBID can be realized by choosing the suitable source to decompose the precursor gas. The organometallic gas used is tungsten carboxyl, W(CO)$_6$, supplied through a nozzle positioned close to the substrate during deposition. The temperature of the precursor is set at 88$^o$ C and the pressure in the chamber ranges between 6-8 $\times$10$^{-6}$ mbar. The scanning of the beam (either electrons or ions) is controlled by a Raith system. We will focus our attention on the EBID structures, which form the primary interest of this work. A schematic of the deposition procedure is shown in Fig. 1a. To optimize the deposition using electron beam, we varied different controllable parameters. The variable beam parameters are the step size, dwell time, electron energy and beam current, in addition to the number of times the scanning is repeated for increased deposition of tungsten. The electron beam scan is carried out using the `line' mode in the Raith system. The substrate used was a doped-Si wafer with a 570 nm thick layer of thermally-grown SiO$_2$ on the surface. 50 nm-thick electrodes of Ti/Au were pre-fabricated lithographically and their edges were etched (using FIB milling) in a wedge-like shape to reduce step-mismatch for the electrical characterization of the EBID nanowires.

\begin{figure}
\begin{center}
\includegraphics[width=85mm]{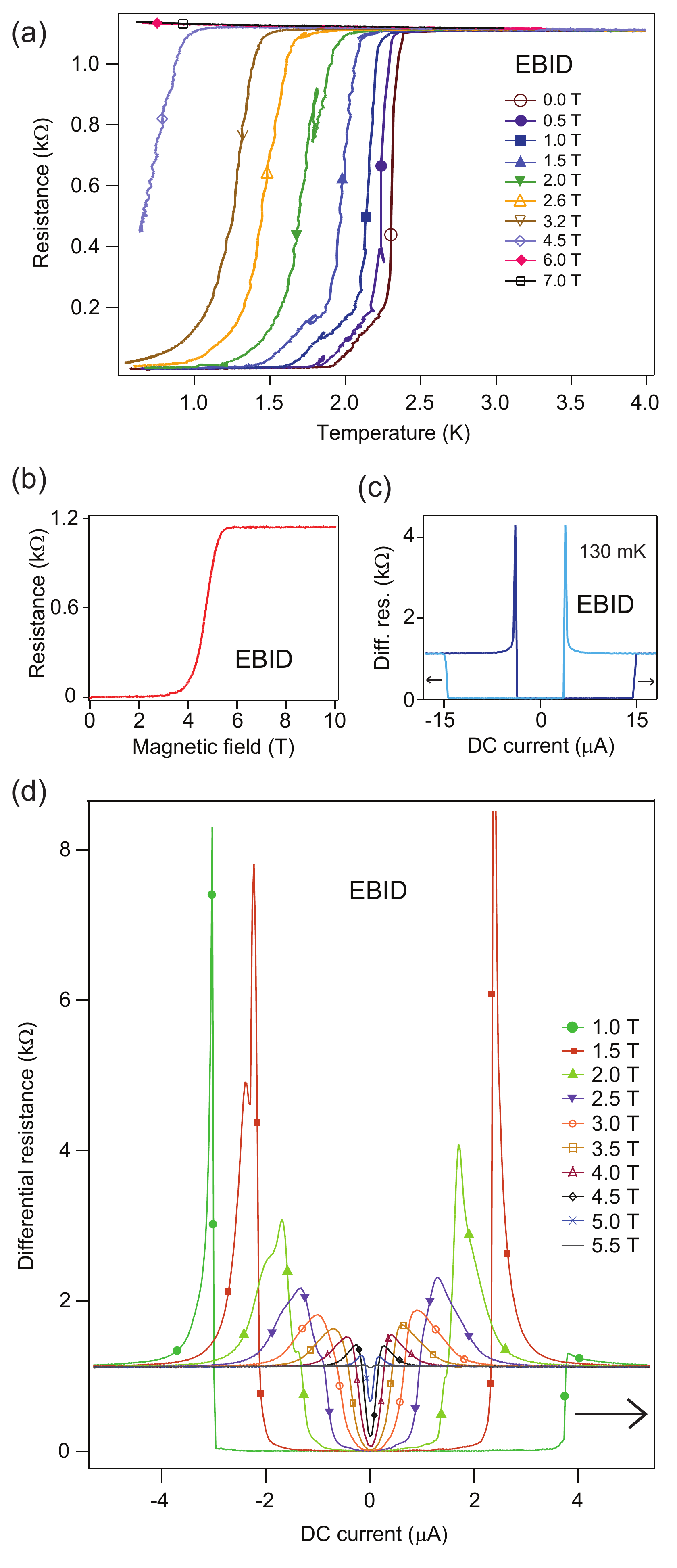}
\caption{(Color online) (a) Resistance as a function of temperature at different magnetic fields for an EBID nanowire. The plot at 0 T suggests a superconducting transition temperature of $T_c$ = 2.0 K. (b) Resistance as a function of magnetic field. (c) Differential resistance ($dV$/$dI$) measurement at zero magnetic field by varying the dc current (I$_{dc}$). The critical current is 14 $\mu$A. (d) Differential resistance ($dV$/$dI$) as a function of dc current (I$_{dc}$) at different magnetic fields. (Note: These are two-probe measurements and a resistance of 80-90 $\Omega$ has been subtracted in the plots to account for the dilution wiring, Ti/Au access leads and possible contact resistance. In figures (b) and (d), temperature varied between 130 mK and 550 mK during the measurements.)}
\end{center}
\end{figure}

When an organometallic gas is decomposed, the resultant product deposited on the substrate consists of amorphous W and C \cite{sadki, ebeam1}. (The W(CO)$_6$ molecule has a W atom bonded to 6 C-O units and it predominantly dissociates into W and CO. Thermodynamic calculations show the possibility of there being some amount of WC also. \cite{thermo}) The parameters of deposition determine the relative percentage of the elements present, drastically affecting the electrical properties of nanowires made by EBID. (Huth et al. \cite{ebeam1} reported 36.9$\%$ W in the best conducting sample. There was no superconductivity and an 8 times reduction in conductance at 2 K as compared to room temperature.) Good deposition is characterized by a strong metallic contrast with respect to the background as seen in the SEM. Our setup included a pair of micro-probes to enable in situ measurement of the resistance of W wires. A wire fabricated with the optimized parameters is shown in Fig. 1b. We used electron energy of 10 keV, step size of 1 nm, dwell time of 10 ms and the scanning (in `line' mode) was repeated 10 times. The beam current was 5.1 nA. The dimension of a typical wire shown in Fig. 1b is 5.9 $\mu$m $\times$ 130 nm $\times$ 120 nm and the room temperature resistance is 1.0 k$\Omega$, yielding a conductivity of 3.7$\times$10$^5$ $\Omega^{-1}$ m$^{-1}$ (an order of magnitude higher than earlier reports \cite{ebeam1, ebeam2}) and a square-resistance of 22 $\Omega$/square. \cite{square} It is expected that a larger dwell time leads to more efficient decomposition of the precursor and hence larger deposition of W. However, we found that with an increased dwell time, the structure obtained has a branched `pillar-like' appearance with upward-growing whiskers instead of a thick continuous wire. (See Fig. 1c.) This structure was obtained using a dwell time of 100 ms (scanning only once in the `line' mode, without any loop repeat). Charging effects on the silicon oxide surface for a very large dwell time causes significant changes to the local electrostatic fields and hinders the deposition of a thick continuous wire, resulting in the `pillar-like' structure instead.

The low temperature electrical properties of the tungsten wires were measured in a dilution refrigerator in a two-probe configuration. The contact pads (which connect the deposited tungsten wire) were wire-bonded to the pins of the sample holder for the measurements. The base temperature of the refrigerator was 130 mK. The temperature of the sample was varied using a heater placed close to it and was read-off with a RuO$_2$ thermometer. The resistance was measured using a standard lock-in technique. The variation of resistance as a function of temperature at different magnetic fields was studied. At zero magnetic field, we observe a sharp drop in the resistance indicating a superconducting transition (Fig. 2a). The superconducting transition is clearly present at higher magnetic fields, at least till 3.2 T as evident from the lineplots in Fig. 2a. We define the critical temperature $T_c$ as the temperature corresponding to a 95$\%$ drop in resistance from its normal state value. At zero magnetic field, this gives $T_c$ = 2.0 K. This set of curves clearly demonstrates that superconducting tungsten wires can be obtained with EBID. \cite{note1}

In earlier studies on EBID tungsten nanowires, Huth et al. \cite{ebeam1} had probed temperatures down to 2 K and Luxmoore et al. \cite{ebeam2} till 1.6 K, but a superconducting transition was not detected. For fabrication of nanowires with EBID, Huth et al. \cite{ebeam1} and Luxmoore et al. \cite{ebeam2} had used a rectangular area raster scan with a pitch of 20 nm between pixels and 100 $\mu$s dwell time. In contrast, we have used a larger dwell time of 10 ms per pixel and smaller step-size of 1 nm in the `line' mode. This may be the key factor for efficient decomposition of the precursor gas and greater amount of tungsten deposition resulting in the largely improved low temperature conductivity of our samples as well as superconductivity.

In Fig. 2b, we plot the magneto-resistance of the tungsten wire. The normal state resistance is restored above 4 T. The critical field is taken to be that magnetic field at which the resistance drops by 95$\%$ of its normal state value. Fig. 2c shows the data for critical current measurement. The differential resistance $dV/dI$ (where $V$ is the voltage drop and $I$ is the current through the device) is measured with a lock-in amplifier by applying a small ac current at a low frequency and recording the voltage drop. The dc current ($I_{dc}$) through the device is continuously varied and its value at which $dV/dI$ switches from zero during the superconductor-to-normal transition is taken to be the critical current $I_c$. There is a hysteretic response in Fig. 2c due to heating in the normal state. The critical current is found to be 14 $\mu$A (consistent for both sweep directions of $I_{dc}$). The measurement of $dV/dI$ as a function of $I_{dc}$ was also done for higher magnetic fields. These are shown in Fig. 2d for different field values. The critical current $I_c$ reduces with an increase in the magnetic field, as expected for a superconductor. (See Fig. 2d.)

\begin{figure}
\begin{center}
\includegraphics[width=65mm]{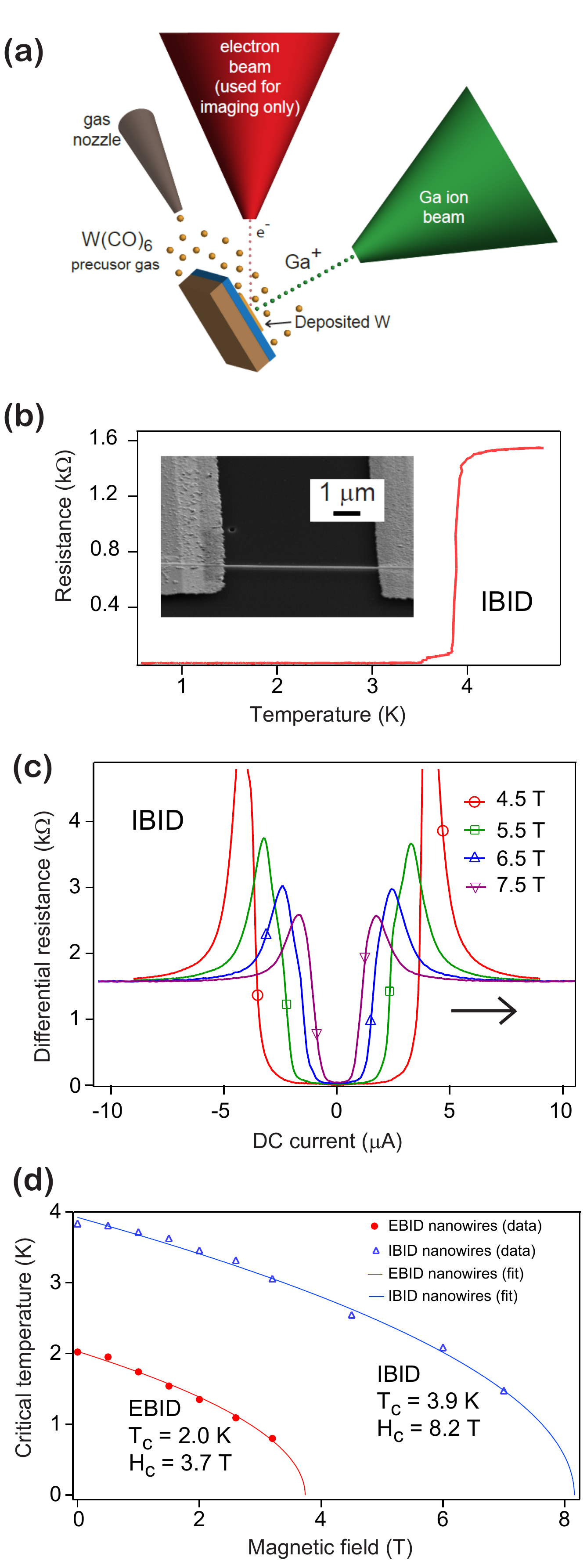}
\caption{(Color online) (a) Schematic of the setup for fabricating IBID tungsten nanowires. The FIB of Ga is used for decomposing the organometallic precursor gas. The electron beam is used for imaging purposes only. (b) Resistance as a function of temperature for an IBID wire. (Inset: SEM image of an IBID wire. The parameters used were: 30 kV FIB energy, 10 pA beam current, 13 nm step size, 0.4 $\mu$s dwell time and a pattern repeat of 100,000 times.) (c) Differential resistance ($dV$/$dI$) measurements as a function of dc current ($I_{dc}$) at different magnetic fields. This shows a critical field greater than 7.5 T. (d) A comparison of the critical temperatures of tungsten wires obtained with EBID and IBID at different magnetic fields. (Note: A resistance of 80-90 $\Omega$ has been subtracted in the plots to account for the dilution wiring, Ti/Au access leads and possible contact resistance. In figure (c), temperature varied between 130 mK and 550 mK during the measurements.)}
\end{center}
\end{figure}

The superconductivity in tungsten is enhanced by a certain degree of amorphousness. Bulk single crystals of tungsten have a critical temperature T$_c$ of only 11 mK. \cite{gibson_bulk_W} But films of $\beta$-W  deposited by electron-bombardment evaporation were seen to have a much higher T$_c$ of 3.35 K. \cite{bond} Enhancement of the critical temperature in evaporated films (as compared to bulk crystals) was observed for some other transition metals like Re and Mo, apart from W. \cite{bond, collver} The work of Collver and Hammond (Ref. \onlinecite{collver}) established the important contribution of disorder in the enhancement of the superconducting critical temperature of tungsten.  These findings support the fact that disordered tungsten nanowires made with EBID and IBID have much higher $T_c$ than bulk crystals. However, it must also be remembered that for these nanowires which contain significant amounts of both W and C, disorder alone is not a sufficient criterion for superconductivity; it is also required that a significant amount of metal (tungsten) be present in the system.

To compare the properties of EBID and IBID wires, we have used our dual-beam deposition system to fabricate tungsten wires with FIB also (Fig. 3a). The result of transport measurements are shown in Fig. 3b. The resistance change indicates a superconducting transition just below 4 K. The critical current for this device was measured to be 48 $\mu$A. Differential resistance measurements are shown in Fig. 3c. The superconductivity is seen to be present for fields as high as 7.5 T. In Fig. 3d, we summarize the critical temperatures of wires made with both EBID at IBID at different magnetic fields. The results are fitted to the empirical relation for the critical temperature T$_{c,H}$ under a finite magnetic field $H$. $\frac{T_{c,H}}{T_c} = \sqrt{1 - \frac{H}{H_c}}$, where $T_c$ is the critical temperature at zero field and $H_c$ is the critical field at absolute zero of temperature. The fit yields $T_c$ = 2.0 K, $H_c$ = 3.7 T for EBID and $T_c$ = 3.9 K, $H_c$ = 8.2 T for IBID nanowires.

EDX (energy-dispersive X-ray spectroscopy) analysis showed a 47$\%\pm$6$\%$ atomic concentration of tungsten in the electron-beam-deposited wire for which transport data has been shown in Fig. 2. The quasi-amorphous nature of the EBID nanowires was confirmed by transmission electron microscope (TEM) images.  For the TEM study, nanowires were fabricated with EBID on carbon membranes using the same deposition parameters discussed earlier in this paper, except that a loop repeat of 5 times was used (instead of 10) for thinner wires. The micro-diffraction pattern shows rings characteristic of nanocrystalline or nanocomposite structure (Fig. 4a). The diameters of the first (most intense) and second rings (measured by the peak positions of intensity) are 4.08 nm$^{-1}$ and 7.29 nm$^{-1}$ respectively (Fig. 4a). These features are qualitatively similar to TEM diffraction images observed by Precht et al. \cite{precht} on disordered tungsten films (deposited by pulsed reactive magnetron sputtering) where signatures of nanocomposite structure were seen on films with 51$\%$ atomic concentration of W. Similar diffraction rings were also seen for our IBID nanowires (fabricated with usual parameters mentioned earlier). (See Fig. 4b). Our findings are compatible with those of Luxmoore et al. \cite{ebeam2} (where intense diffraction rings were seen only for superconducting IBID nanowires, but broad and weak reflections were seen in the case of non-superconducting EBID nanowires), and points to a correlation between the structure of the nanowires (like nanocomposite or short-range nano-crystalline order) and the presence of superconducting behaviour.

In extending this work of realizing superconducting EBID tungsten wires to practical devices, the issue of contamination needs to be addressed. During the deposition of EBID wires, there are carbon compounds formed by decomposition of the organometallic gas which are deposited and can extend to 1 $\mu$m on either side (shown by dotted lines in Fig. 1b). This can pose a problem if EBID deposition is carried out for making contacts to nanomaterials with a channel length of less than 2 $\mu$m, since the contamination may be deposited on the device and mask its intrinsic transport properties. Protecting the device by a resist or a metal mask during the deposition may be required in such cases.

\begin{figure}
\begin{center}
\includegraphics[width=85mm]{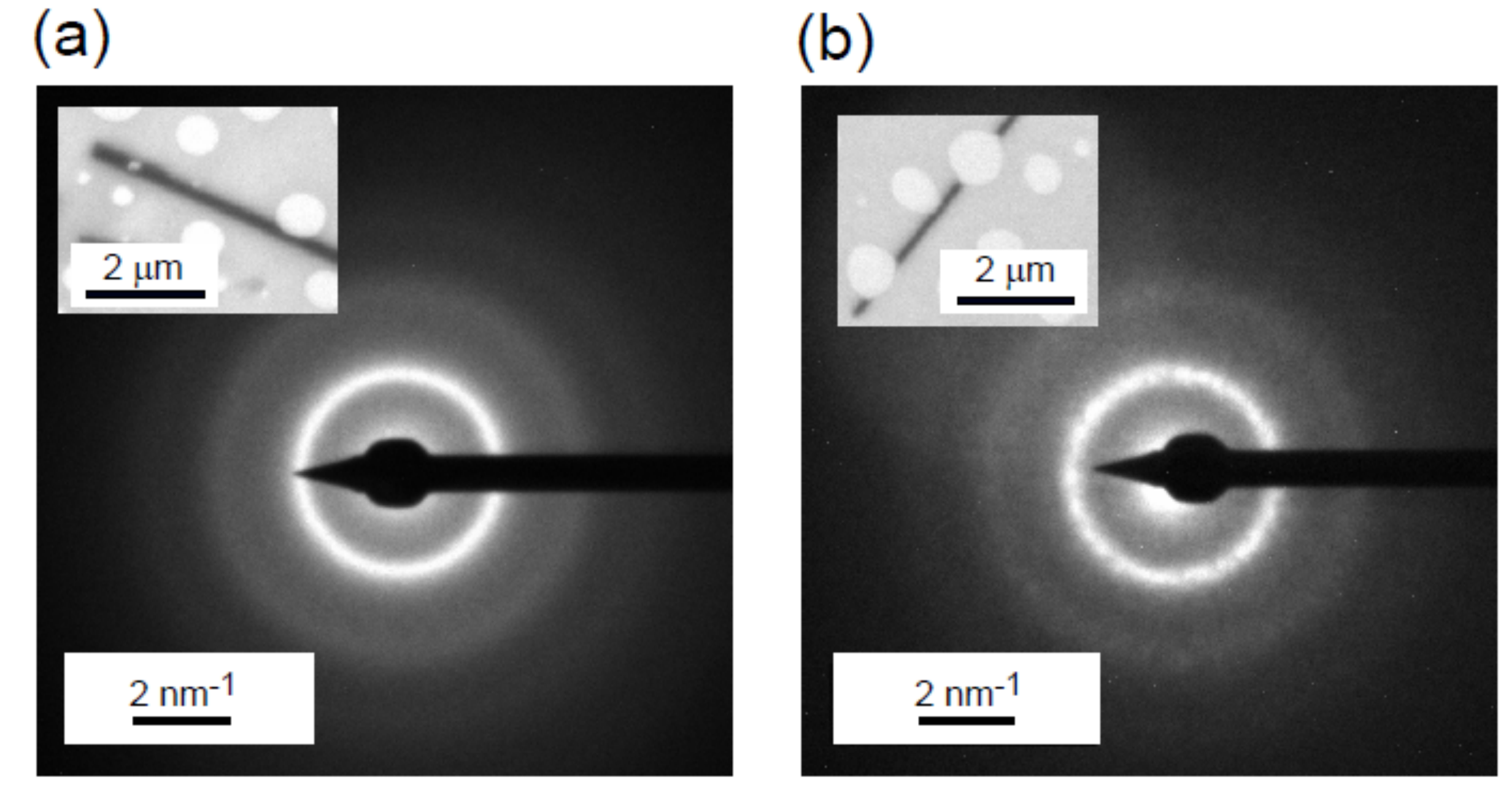}
\caption{(a) TEM micro-diffraction pattern of an EBID nanowire (inset). (b) TEM micro-diffraction pattern for an IBID nanowire (inset).}
\end{center}
\end{figure}

In this work, we have demonstrated that tungsten nanowires fabricated using electron beam induced decomposition of an organometallic gas undergo a superconducting transition. This solves a long-standing problem that EBID wires were not only non-superconducting but also non-metallic at low temperatures. EBID is less destructive than the conventional method of IBID and it is a major advantage of this process. EBID has found a wide range of applications like rapid micro-scale patterning of nanostructures \cite{cobalt, dot_array}, fabrication of DNA tunneling detectors \cite{dna} and experiments with plasmonic nanoclusters \cite{plasmonic}. Our results add to this list and can show new directions of research concerning superconducting devices, disordered superconductivity, devices based on the physics of such systems and experiments of proximity-induced superconductivity in fragile nanowires and ultrathin materials. 

We thank Odile Kaitasov, Mandar Deshmukh and Richard Deblock for discussions. This work was supported by ANR Supergraph.


\begin{thebibliography}{99}
\bibitem{square}
For a nanowire of rectangular cross-section with length $L$ and width $w$, the resistance $R$ and square-resistance $R_s$ are related by $R_s = R/(L/w)$.
\bibitem{note1}
The superconductor - normal metal transition in Fig. 2a at low magnetic field shows a `knee' followed by a sharp increase in the resistance. It is possible that this knee-like feature originates from the transition of the contact region of the nanowire, which might have a lower $T_c$ than the rest of its length.
\end{thebibliography}

\end{document}